\documentclass[aps,prl,twocolumn,superscriptaddress,preprintnumbers,amsmath,amssymb]{revtex4}
\usepackage{graphicx,color,dcolumn,booktabs,bm}
\usepackage{longtable,lscape}
\usepackage{amsmath}
\usepackage{indentfirst}
\usepackage{epsfig}
\usepackage{feynmf}   %{feynmp}
\usepackage{epstopdf}   %{feynmp}
\usepackage{slashed}  %for Feynman symbols
\usepackage{cases}
\usepackage{color}
\usepackage{multirow}
\usepackage{graphicx,color,dcolumn,booktabs,bm}
\usepackage{verbatim}
\usepackage[colorlinks,linkcolor=red,anchorcolor=green,citecolor=blue]{hyperref}
\usepackage{mathrsfs}
\usepackage{rotating}
\usepackage{threeparttable}
\usepackage{lineno}
\newcommand{\add}[1]{{\color{blue}{#1}}}

\newcommand{\BR}{{\cal B}}

\newcommand{\beq}{\begin{equation}}
\newcommand{\eeq}{\end{equation}}
\newcommand{\bitm}{\begin{itemize}}
\newcommand{\eitm}{\end{itemize}}

\begin{document}
%\title{\quad\\[1.0cm] Search for a light Higgs boson in single-photon decays of \boldmath{$\Upsilon(1S)$} using \boldmath{$\Upsilon(2S) \to \pi^+ \pi^- \Upsilon(1S)$} tagging method}
\title{Search for a light Higgs boson in single-photon decays of \boldmath{$\Upsilon(1S)$} using \boldmath{$\Upsilon(2S) \to \pi^+ \pi^- \Upsilon(1S)$} tagging method}

\noaffiliation
\affiliation{Department of Physics, University of the Basque Country UPV/EHU, 48080 Bilbao}
\affiliation{University of Bonn, 53115 Bonn}
\affiliation{Brookhaven National Laboratory, Upton, New York 11973}
\affiliation{Budker Institute of Nuclear Physics SB RAS, Novosibirsk 630090}
\affiliation{Faculty of Mathematics and Physics, Charles University, 121 16 Prague}
%%%\affiliation{Chiba University, Chiba 263-8522}
\affiliation{Chonnam National University, Gwangju 61186}
\affiliation{Chung-Ang University, Seoul 06974}
\affiliation{University of Cincinnati, Cincinnati, Ohio 45221}
\affiliation{Deutsches Elektronen--Synchrotron, 22607 Hamburg}
\affiliation{Duke University, Durham, North Carolina 27708}
\affiliation{Institute of Theoretical and Applied Research (ITAR), Duy Tan University, Hanoi 100000}
\affiliation{University of Florida, Gainesville, Florida 32611}
\affiliation{Department of Physics, Fu Jen Catholic University, Taipei 24205}
\affiliation{Key Laboratory of Nuclear Physics and Ion-beam Application (MOE) and Institute of Modern Physics, Fudan University, Shanghai 200443}
%%%\affiliation{Justus-Liebig-Universit\"at Gie\ss{}en, 35392 Gie\ss{}en}
\affiliation{Gifu University, Gifu 501-1193}
%%%\affiliation{II. Physikalisches Institut, Georg-August-Universit\"at G\"ottingen, 37073 G\"ottingen}
\affiliation{SOKENDAI (The Graduate University for Advanced Studies), Hayama 240-0193}
\affiliation{Gyeongsang National University, Jinju 52828}
\affiliation{Department of Physics and Institute of Natural Sciences, Hanyang University, Seoul 04763}
\affiliation{University of Hawaii, Honolulu, Hawaii 96822}
\affiliation{High Energy Accelerator Research Organization (KEK), Tsukuba 305-0801}
\affiliation{J-PARC Branch, KEK Theory Center, High Energy Accelerator Research Organization (KEK), Tsukuba 305-0801}
\affiliation{National Research University Higher School of Economics, Moscow 101000}
\affiliation{Forschungszentrum J\"{u}lich, 52425 J\"{u}lich}
%%%\affiliation{Hiroshima University, Higashi-Hiroshima, Hiroshima 739-8530}
\affiliation{IKERBASQUE, Basque Foundation for Science, 48013 Bilbao}
\affiliation{Indian Institute of Science Education and Research Mohali, SAS Nagar, 140306}
%%%\affiliation{Indian Institute of Technology Bhubaneswar, Satya Nagar 751007}
\affiliation{Indian Institute of Technology Guwahati, Assam 781039}
\affiliation{Indian Institute of Technology Hyderabad, Telangana 502285}
\affiliation{Indian Institute of Technology Madras, Chennai 600036}
\affiliation{Indiana University, Bloomington, Indiana 47408}
%%%\affiliation{Institute of High Energy Physics, Chinese Academy of Sciences, Beijing 100049}
\affiliation{Institute of High Energy Physics, Vienna 1050}
\affiliation{Institute for High Energy Physics, Protvino 142281}
%%%\affiliation{Institute of Mathematical Sciences, Chennai 600113}
\affiliation{INFN - Sezione di Napoli, I-80126 Napoli}
\affiliation{INFN - Sezione di Roma Tre, I-00146 Roma}
\affiliation{INFN - Sezione di Torino, I-10125 Torino}
\affiliation{Iowa State University, Ames, Iowa 50011}
\affiliation{Advanced Science Research Center, Japan Atomic Energy Agency, Naka 319-1195}
\affiliation{J. Stefan Institute, 1000 Ljubljana}
%%%\affiliation{Kanagawa University, Yokohama 221-8686}
\affiliation{Institut f\"ur Experimentelle Teilchenphysik, Karlsruher Institut f\"ur Technologie, 76131 Karlsruhe}
\affiliation{Kavli Institute for the Physics and Mathematics of the Universe (WPI), University of Tokyo, Kashiwa 277-8583}
%%%\affiliation{King Abdulaziz City for Science and Technology, Riyadh 11442}
\affiliation{Department of Physics, Faculty of Science, King Abdulaziz University, Jeddah 21589}
\affiliation{Kitasato University, Sagamihara 252-0373}
\affiliation{Korea Institute of Science and Technology Information, Daejeon 34141}
\affiliation{Korea University, Seoul 02841}
%%%\affiliation{Kyoto Sangyo University, Kyoto 603-8555}
\affiliation{Kyungpook National University, Daegu 41566}
\affiliation{Universit\'{e} Paris-Saclay, CNRS/IN2P3, IJCLab, 91405 Orsay}
%%%\affiliation{\'Ecole Polytechnique F\'ed\'erale de Lausanne (EPFL), Lausanne 1015}
\affiliation{P.N. Lebedev Physical Institute of the Russian Academy of Sciences, Moscow 119991}
%%%\affiliation{Liaoning Normal University, Dalian 116029}
\affiliation{Faculty of Mathematics and Physics, University of Ljubljana, 1000 Ljubljana}
\affiliation{Ludwig Maximilians University, 80539 Munich}
\affiliation{Luther College, Decorah, Iowa 52101}
\affiliation{Malaviya National Institute of Technology Jaipur, Jaipur 302017}
%%%\affiliation{University of Malaya, 50603 Kuala Lumpur}
\affiliation{Faculty of Chemistry and Chemical Engineering, University of Maribor, 2000 Maribor}
\affiliation{Max-Planck-Institut f\"ur Physik, 80805 M\"unchen}
\affiliation{School of Physics, University of Melbourne, Victoria 3010}
\affiliation{University of Mississippi, University, Mississippi 38677}
\affiliation{University of Miyazaki, Miyazaki 889-2192}
\affiliation{Moscow Physical Engineering Institute, Moscow 115409}
\affiliation{Graduate School of Science, Nagoya University, Nagoya 464-8602}
%%%\affiliation{Kobayashi-Maskawa Institute, Nagoya University, Nagoya 464-8602}
\affiliation{Universit\`{a} di Napoli Federico II, I-80126 Napoli}
\affiliation{Nara Women's University, Nara 630-8506}
\affiliation{National Central University, Chung-li 32054}
%%%\affiliation{National United University, Miao Li 36003}
\affiliation{Department of Physics, National Taiwan University, Taipei 10617}
\affiliation{H. Niewodniczanski Institute of Nuclear Physics, Krakow 31-342}
\affiliation{Nippon Dental University, Niigata 951-8580}
\affiliation{Niigata University, Niigata 950-2181}
\affiliation{University of Nova Gorica, 5000 Nova Gorica}
\affiliation{Novosibirsk State University, Novosibirsk 630090}
%%%\affiliation{Okinawa Institute of Science and Technology, Okinawa 904-0495}
\affiliation{Osaka City University, Osaka 558-8585}
\affiliation{Pacific Northwest National Laboratory, Richland, Washington 99352}
\affiliation{Panjab University, Chandigarh 160014}
%%%\affiliation{Peking University, Beijing 100871}
\affiliation{University of Pittsburgh, Pittsburgh, Pennsylvania 15260}
\affiliation{Punjab Agricultural University, Ludhiana 141004}
%%%\affiliation{Research Center for Electron Photon Science, Tohoku University, Sendai 980-8578}
\affiliation{Research Center for Nuclear Physics, Osaka University, Osaka 567-0047}
\affiliation{Meson Science Laboratory, Cluster for Pioneering Research, RIKEN, Saitama 351-0198}
%%%\affiliation{Theoretical Research Division, Nishina Center, RIKEN, Saitama 351-0198}
%%%\affiliation{RIKEN BNL Research Center, Upton, New York 11973}
\affiliation{Dipartimento di Matematica e Fisica, Universit\`{a} di Roma Tre, I-00146 Roma}
\affiliation{Department of Modern Physics and State Key Laboratory of Particle Detection and Electronics, University of Science and Technology of China, Hefei 230026}
%%%\affiliation{Seoul National University, Seoul 08826}
\affiliation{Showa Pharmaceutical University, Tokyo 194-8543}
%%%\affiliation{Soochow University, Suzhou 215006}
\affiliation{Soongsil University, Seoul 06978}
%%%\affiliation{University of South Carolina, Columbia, South Carolina 29208}
%%%\affiliation{Stefan Meyer Institute for Subatomic Physics, Vienna 1090}
\affiliation{Sungkyunkwan University, Suwon 16419}
\affiliation{School of Physics, University of Sydney, New South Wales 2006}
\affiliation{Department of Physics, Faculty of Science, University of Tabuk, Tabuk 71451}
\affiliation{Tata Institute of Fundamental Research, Mumbai 400005}
%%%\affiliation{Excellence Cluster Universe, Technische Universit\"at M\"unchen, 85748 Garching}
\affiliation{Department of Physics, Technische Universit\"at M\"unchen, 85748 Garching}
%%%\affiliation{School of Physics and Astronomy, Tel Aviv University, Tel Aviv 69978}
\affiliation{Toho University, Funabashi 274-8510}
\affiliation{Department of Physics, Tohoku University, Sendai 980-8578}
\affiliation{Earthquake Research Institute, University of Tokyo, Tokyo 113-0032}
\affiliation{Department of Physics, University of Tokyo, Tokyo 113-0033}
\affiliation{Tokyo Institute of Technology, Tokyo 152-8550}
\affiliation{Tokyo Metropolitan University, Tokyo 192-0397}
%%%\affiliation{Utkal University, Bhubaneswar 751004}
\affiliation{Virginia Polytechnic Institute and State University, Blacksburg, Virginia 24061}
\affiliation{Wayne State University, Detroit, Michigan 48202}
\affiliation{Yamagata University, Yamagata 990-8560}
\affiliation{Yonsei University, Seoul 03722}

  \author{S.~Jia}\affiliation{Key Laboratory of Nuclear Physics and Ion-beam Application (MOE) and Institute of Modern Physics, Fudan University, Shanghai 200443} % Fudan
  \author{C.~P.~Shen}\affiliation{Key Laboratory of Nuclear Physics and Ion-beam Application (MOE) and Institute of Modern Physics, Fudan University, Shanghai 200443} % Fudan
  \author{I.~Adachi}\affiliation{High Energy Accelerator Research Organization (KEK), Tsukuba 305-0801}\affiliation{SOKENDAI (The Graduate University for Advanced Studies), Hayama 240-0193} % KEK
% \author{K.~Adamczyk}\affiliation{H. Niewodniczanski Institute of Nuclear Physics, Krakow 31-342} % Krakow
% \author{J.~K.~Ahn}\affiliation{Korea University, Seoul 02841} % Korea
  \author{H.~Aihara}\affiliation{Department of Physics, University of Tokyo, Tokyo 113-0033} % Tokyo
  \author{S.~Al~Said}\affiliation{Department of Physics, Faculty of Science, University of Tabuk, Tabuk 71451}\affiliation{Department of Physics, Faculty of Science, King Abdulaziz University, Jeddah 21589} % Tabuk
  \author{D.~M.~Asner}\affiliation{Brookhaven National Laboratory, Upton, New York 11973} % BNL
  \author{H.~Atmacan}\affiliation{University of Cincinnati, Cincinnati, Ohio 45221} % Cincinnati
% \author{V.~Aulchenko}\affiliation{Budker Institute of Nuclear Physics SB RAS, Novosibirsk 630090}\affiliation{Novosibirsk State University, Novosibirsk 630090} % BINP
  \author{T.~Aushev}\affiliation{National Research University Higher School of Economics, Moscow 101000} % HSE
  \author{R.~Ayad}\affiliation{Department of Physics, Faculty of Science, University of Tabuk, Tabuk 71451} % Tabuk
% \author{T.~Aziz}\affiliation{Tata Institute of Fundamental Research, Mumbai 400005} % Tata
  \author{V.~Babu}\affiliation{Deutsches Elektronen--Synchrotron, 22607 Hamburg} % DESY
% \author{S.~Bahinipati}\affiliation{Indian Institute of Technology Bhubaneswar, Satya Nagar 751007} % IITB
% \author{A.~M.~Bakich}\affiliation{School of Physics, University of Sydney, New South Wales 2006} % Sydney
% \author{Y.~Ban}\affiliation{Peking University, Beijing 100871} % Peking
% \author{E.~Barberio}\affiliation{School of Physics, University of Melbourne, Victoria 3010} % Melbourne
% \author{M.~Barrett}\affiliation{High Energy Accelerator Research Organization (KEK), Tsukuba 305-0801} % KEK
% \author{M.~Bauer}\affiliation{Institut f\"ur Experimentelle Teilchenphysik, Karlsruher Institut f\"ur Technologie, 76131 Karlsruhe} % Karlsruhe
% ...
% Lebedev
% \author{V.~Zhulanov}\affiliation{Budker Institute of Nuclear Physics SB RAS, Novosibirsk 630090}\affiliation{Novosibirsk State University, Novosibirsk 630090} % BINP
\collaboration{The Belle Collaboration}

\begin{abstract}

%\linenumbers

We search for a light Higgs boson ($A^0$) decaying into a $\tau^+\tau^-$ or $\mu^+\mu^-$ pair in the radiative decays of $\Upsilon(1S)$.
The production of $\Upsilon(1S)$ mesons is tagged by $\Upsilon(2S)\to \pi^+ \pi^- \Upsilon(1S)$ transitions, using 158 million $\Upsilon(2S)$ events accumulated with the Belle detector at the KEKB asymmetric energy electron-positron collider. No significant $A^0$ signals in the mass range from the $\tau^+\tau^-$ or $\mu^+\mu^-$ threshold to 9.2 GeV/$c^2$ are observed. We set the upper limits at 90\% credibility level (C.L.) on the product branching fractions for $\Upsilon(1S)\to \gamma A^0$ and $A^0\to \tau^+\tau^-$ varying from $3.8\times10^{-6}$ to $1.5\times10^{-4}$.
Our results represent an approximately twofold improvement on the current world best upper limits for the $\Upsilon(1S)\to \gamma A^0(\to \tau^+\tau^-)$ production. For $A^0\to \mu^+\mu^-$, the upper limits on the product branching fractions for $\Upsilon(1S)\to \gamma A^0$ and $A^0\to \mu^+\mu^-$ are at the same level as the world average limits, and vary from $3.1\times10^{-7}$ to $1.6\times10^{-5}$.
The upper limits at 90\% C.L. on the Yukawa coupling $f_{\Upsilon(1S)}$ and mixing angle ${\rm sin}\theta_{A^0}$ are also given.

\end{abstract}

%\pacs{13.66.Bc, 13.87.Fh, 14.40.Lb}

\maketitle

%\linenumbers
%%%%%%%%%%%%%%%%%%%%%%%%%%%%%%%%%%%%%%%%%%%%%%%%%%%%%%%%%%%%%%%%%%%%%%%

In 2012, the last missing Standard Model (SM) particle, a Higgs boson, was discovered by ATLAS and CMS~\cite{7161,71630}, demonstrating that the Higgs mechanism would break the electroweak symmetry and give rise to the masses of $W$ and $Z$ bosons as well as quarks and leptons~\cite{132,508}. Besides this massive Higgs boson, three $CP$-even, two $CP$-odd, and two charged Higgs bosons are predicted by the Next-to Minimal Supersymmetric Standard Model (NMSSM)~\cite{034018,041801,051105,015013,075003}. NMSSM adds an additional singlet chiral superfield to the Minimal Supersymmetric Standard Model (MSSM)~\cite{117} to address the so-called ``little hierarchy problem"~\cite{710}, in which the value of the supersymmetric Higgs mass parameter $\mu$ is many orders of magnitude below the Planck scale.

The lightest $CP$-odd Higgs boson, denoted as $A^0$, could have a mass smaller than twice the mass of the $b$ quark, making it accessible via radiative $\Upsilon(nS) \to \gamma A^0$ ($n$ = 1, 2, 3) decays~\cite{034018,041801,051105,015013,075003,1304}. The coupling of the $A^0$ to $\tau^+\tau^-$ and $b\bar b$ is proportional to ${\rm tan}\beta{\rm cos}\theta_{A^0}$, where ${\rm tan}\beta$ is the ratio of vacuum expectation values for the two Higgs doublets, and $\theta_{A^0}$ is the mixing angle between doublet and singlet $CP$-odd Higgs bosons~\cite{051105}. The branching fraction of $\Upsilon(nS) \to \gamma A^0$ could be as large as $10^{-4}$, depending on the values of the $A^0$ mass, ${\rm tan}\beta$, and ${\rm cos}\theta_A$~\cite{051105}. For $2m_\tau < m_{A^0} < 2m_b$, the decay of $A^0\to \tau^+\tau^-$ is expected to dominate~\cite{051105,015018}. For $m_{A^0} < 2m_\tau$, the $A^0 \to \mu^+\mu^-$ events can be copiously produced~\cite{015018}.

Identifying the origin and nature of dark matter (DM) is a longstanding unsolved problem in astronomy and particle physics.~One type of DM, often called the weakly interacting massive particle (WIMP), is generally expected to be in the mass region ranging from ${\cal O}$(1) MeV~\cite{279,101301} to ${\cal O}$(100) TeV~\cite{686162,39,119,157,039,1609162}. An extensive experimental search program has been devoted to WIMPs with the electroweak mass, but no clear evidence has been found to date~\cite{842}. In recent years, the possibility that WIMPs  have a mass at or below the GeV-scale has gained much attraction.
%\add{For example, the decay of $\Upsilon(nS) \to \gamma H$ followed by the $H$ decaying into DM particles $\chi$ and lepton pairs such as $\tau^+\tau^-$ and $\mu^+\mu^-$ is suggested for to be searched for in the $B$-factories~\cite{054034,115019,042,050,03553}, where $H$ is the mediator having an interaction between the WIMP and SM particles.
%BaBar and Belle have searched for the on-shell process $\Upsilon(1S)\to\gamma A^0$ with $A^0\to \chi\chi$ and off-shell process $\Upsilon(1S)\to\gamma\chi\chi$. No
%evidence for a signal was found, and upper limits on the branching fractions of such processes were set~\cite{021804,011801}.
For example, the decay of $\Upsilon(nS) \to \gamma H$ followed by the $H$ decaying into a lepton pair such as $\tau^+\tau^-$ and $\mu^+\mu^-$ is suggested to be searched for in the $B$-factories~\cite{042,050,03553}, where $H$ is the mediator having an interaction between the WIMP and SM particles.

BaBar and Belle have searched for $A^0$ decaying into a pair of low mass dark matter with the invisible final-states in $\Upsilon(1S)$ radiative decays~\cite{021804,011801}. Searches for $A^0$ decaying into $\tau^+\tau^-$ and $\mu^+\mu^-$ have been also performed in $\Upsilon(1S,2S,3S)$ radiative decays by CLEO~\cite{151802} and BaBar~\cite{071102,031102,081803,181801}. No significant signals were found. The upper limits at 90\% C.L. on the product of branching fractions $\BR(\Upsilon(nS)\to\gamma A^0)\BR(A^0 \to \tau^+\tau^-/\mu^+\mu^-)$ ($n$ = 1, 2, 3) have been set at levels of $10^{-6}$ and $10^{-5}$. In particular, for $\Upsilon(1S)$ decays, more stringent upper limits are obtained by BaBar~\cite{071102,031102}.
%The upper limits at 90\% C.L. on $\BR(\Upsilon(1S)\to\gamma A^0)\BR(A^0 \to \tau^+\tau^-)$ vary from $0.9\times10^{-5}$ to $13\times10^{-5}$ with $A^0$ masses ranging from $2m_{\tau}$ to 9.2 GeV/$c^2$~\cite{071102}, and on $\BR(\Upsilon(1S)\to\gamma A^0)\BR(A^0 \to \mu^+\mu^-)$ vary from $0.28\times10^{-6}$ to $9.7\times10^{-6}$ with $A^0$ masses ranging from $2m_{\mu}$ to 9.2 GeV/$c^2$~\cite{031102}.

In this Letter, we conduct a search for the light $CP$-odd Higgs boson $A^0$ in $\Upsilon(1S)$ radiative decays with $A^0\to \tau^+\tau^-$ and $A^0\to \mu^+\mu^-$. This search is based on an $\Upsilon(2S)$ data sample with the integrated luminosity of 24.91 fb$^{-1}$, corresponding to $(158\pm4)\times10^6$ $\Upsilon(2S)$ events, collected by the Belle detector~\cite{Belle1} at the KEKB asymmetric-energy $e^+e^-$ collider~\cite{KEKB1}.
A detailed description of the Belle detector can be found in Refs.~\cite{Belle1}.
The $\Upsilon(1S)$ mesons are selected via the $\Upsilon(2S)\to \pi^+\pi^- \Upsilon(1S)$ transitions. In this case one must trigger and reconstruct final states in which two extra low momentum pions are identified in the detector, trying to avoid collecting too many background events and at the same time maintaining a high trigger efficiency. We assume that the width of $A^0$ can be neglected compared to the experimental resolution and the lifetime of $A^0$ is short enough~\cite{65367}.

We use {\sc evtgen}~\cite{462152} to generate signal Monte Carlo (MC) events to determine signal line shapes and efficiencies, and optimize selection criteria. The VVPIPI model~\cite{462152} is used to generate the decay $\Upsilon(2S)\to \pi^+\pi^-\Upsilon(1S)$. The angle of the radiative photon in the $\Upsilon(1S)$ frame ($\theta_\gamma$) is distributed according to $1 + {\rm cos}^2\theta_\gamma$ for $\Upsilon(1S)\to \gamma A^0$. The effect of final-state radiation (FSR) is taken into account in the simulation using the PHOTOS package~\cite{291}.
The simulated events are processed with a detector simulation based on {\sc geant3}~\cite{geant3}.
Multiple $A^0$ masses are generated: 3.6(0.22) GeV/$c^2$ to 9.2 GeV/$c^2$ in steps of 0.5 GeV/$c^2$ or less for $A^0\to \tau^+\tau^-(\mu^+\mu^-)$. Inclusive MC samples of $\Upsilon(2S)$ decays with four times the luminosity as the real data are produced to check possible peaking backgrounds from $\Upsilon(2S)$ decays~\cite{107540}.

The entire decay channel can be written as $\Upsilon(2S)\to \pi^+\pi^- \Upsilon(1S)$, $\Upsilon(1S)\to \gamma A^0$, $A^0\to \tau^+\tau^-/\mu^+\mu^-$. In selecting $A^0\to \tau^+\tau^-$ candidates, at least one tau lepton decays leptonically, resulting in five different combinations: $\tau\tau\to$ $ee$, $\mu\mu$, $e\mu$, $e\pi$, $\mu\pi$, writing with neutrinos omitted. Note that $\tau^-\to \pi^-\nu_{\tau}$, $\tau^-\to \pi^-\nu_{\tau}+n\pi^0$ ($n\ge 1$), are all included in $\tau\to\pi$. Events in which both tau leptons decay hadronically ($\tau\tau\to\pi\pi$) \add{suﬀer} from significantly larger and poorly modeled backgrounds than in the leptonic channels, and therefore this mode is excluded.

The charged tracks and particle identifications for the pions and leptons are performed using the same method as in Ref.~\cite{121803}.
An electromagnetic calorimeter cluster is treated as a photon candidate if it is isolated from the projected path of charged tracks in the central drift chamber. The energy of photons is required to be larger than 50 MeV. The most energetic photon is regarded as the $\Upsilon(1S)$ radiative photon.

For $A^0\to \tau^+\tau^-$, the missing energy in the laboratory frame is required to be greater than 2 GeV to suppress non-$\tau$ decays and ISR backgrounds. The dominant backgrounds come from $\Upsilon(2S)\to\pi^+\pi^-\Upsilon(1S)(\to \ell^+\ell^-(\gamma))$ ($\ell$ = $e$, $\mu$, or $\tau$) decays, which have an event topology similar to that of the signal.~The backgrounds from $\pi^0$ decays are also large, where photons from $\pi^0$ decays are misidentified as $\Upsilon(1S)$ radiative photons, especially when the energy of $\Upsilon(1S)$ radiative photon is low. To reduce such backgrounds, a likelihood function is employed to distinguish isolated photons from $\pi^0$ daughters using the invariant mass of the photon pair, photon energy in the laboratory frame, and the angle with respect to the beam direction in the laboratory frame~\cite{061803}. We combine the signal photon candidate with any other photon and then reject both photons of a pair whose $\pi^0$ likelihood is larger than 0.3. To further suppress $\pi^0$ backgrounds in $\rho^{\pm}\to \pi^{\pm}\pi^0$, we require cos$\theta(\gamma \pi^{\pm})$ $<$ 0.4, where cos$\theta(\gamma \pi^{\pm})$ is the cosine of the angle between the photon from $\Upsilon(1S)$ decays and $\pi^{\pm}$ from $\tau^{\pm}$ decays in the laboratory frame. We impose requirements of cos$\theta(\gamma e)$ $<$ 0.95 and cos$\theta(\gamma \mu)$ $<$ 0.8 to remove FSR and $\Upsilon(1S) \to \mu^+ \mu^-(\gamma)/e^+e^-(\gamma)$ backgrounds, where cos$\theta(\gamma e)$ and cos$\theta(\gamma \mu)$ are the cosine of the angle between the $\Upsilon(1S)$ radiative photon and $e$ and $\mu$ from $\tau$ decays in the laboratory frame. All of the above selection criteria have been optimized by maximizing FOM = $N_{\rm sig}/\sqrt{N_{\rm sig}+N_{\rm bkg}}$, where $N_{\rm sig}$ is the expected signal yield from signal MC samples assuming $\BR(\Upsilon(1S)\to \gamma A^0)\BR(A^0\to \tau^+\tau^-)$ = 10$^{-5}$~\cite{151802,071102}, and $N_{\rm bkg}$ is the number of normalized background events from inclusive MC samples.

For $A^0\to \mu^+\mu^-$, a four-constraint (4C) kinematic fit constraining the four-momenta of the final-state particles to the initial $e^+e^-$ collision system is performed to suppress backgrounds with multiple photons and improve mass resolutions. The $\chi^2/{\rm ndf}$ of the 4C fit is required to be less than 12.5, where the number of degrees of freedom (ndf) is four. The cosine of the angle between the $\Upsilon(1S)$ radiative photon and $\mu$ is required to be less than 0.8 to suppress FSR and $\Upsilon(1S) \to \mu^+ \mu^-(\gamma)$ backgrounds. These requirements have also been optimized using the FOM method assuming $\BR(\Upsilon(1S)\to \gamma A^0)\BR(A^0\to \mu^+\mu^-) = 10^{-6}$~\cite{151802,031102}.

The $\Upsilon(1S)$ is tagged by the requirement on the mass recoiling against a pion pair (recoil mass). The best candidate is chosen by selecting the recoil mass of dipion closest to the $\Upsilon(1S)$ nominal mass~\cite{PDG}.

\begin{figure}[htbp]
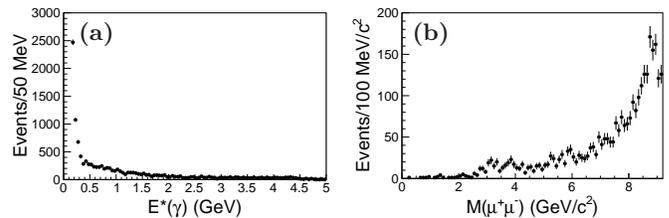

\centering
\includegraphics[width=2.78cm,angle=-90]{fig1a.epsi}
\put(-95, -12){\bf (a)}
\hspace{0.2cm}
\includegraphics[width=2.78cm,angle=-90]{fig1b.epsi}
\put(-95, -12){\bf (b)}
\caption{The (a) $E^*(\gamma)$ and (b) $M(\mu^+\mu^-)$ distributions from the $\Upsilon(2S)$ data sample.}\label{taumu}
\end{figure}

Considering $\tau$ decays with undetected neutrinos, we identify the $A^0$ signal using the photon energy in the $\Upsilon(1S)$ rest frame ($E^*(\gamma)$), which can be converted to $M(\tau^+\tau^-)$ via $M^2(\tau^+\tau^-)$ = $m^2_{\Upsilon(1S)}-2m_{\Upsilon(1S)}E^*(\gamma)$, where $m_{\Upsilon(1S)}$ is the nominal mass of $\Upsilon(1S)$~\cite{PDG}. Hereinafter, $M$ represents a measured invariant mass. For $A^0\to \mu^+\mu^-$, we identify the $A^0$ signal using the invariant mass distribution of $\mu^+\mu^-$ ($M(\mu^+\mu^-)$). After requiring the events within the $\Upsilon(1S)$ signal region of [9.45, 9.47] GeV/$c^2$ and the application of the above requirements, the $E^*(\gamma)$ and $M(\mu^+\mu^-)$ distributions from the $\Upsilon(2S)$ data sample are as shown in Fig.~\ref{taumu}. No significant signals are seen.

For $A^0\to\tau^+\tau^-$, we perform a series of two-dimensional (2D) unbinned maximum-likelihood fits to $E^*(\gamma)$ and $M_{\rm rec}(\pi^+\pi^-)$ distributions to extract the $\Upsilon(1S)\to\gamma A^0(\to \tau^+\tau^-)$ signal yields. The 2D fitting function $f(E,M)$ is expressed as
\begin{equation} \label{eq:q1}
\begin{aligned}
f(E,M) = N^{\rm sig}s_1(E)s_2(M)+N^{\rm bg}_{\rm sb}s_1(E)b_2(M)+\\
N^{\rm bg}_{\rm bs}b_1(E)s_2(M)+N^{\rm bg}_{\rm bb}b_1(E)b_2(M),
\end{aligned}
\end{equation}
where $s_1(E)$ and $b_1(E)$ are the signal and background probability density functions (PDFs) for the $E^{*}(\gamma)$ distributions, and $s_2(M)$ and $b_2(M)$ are the corresponding PDFs for the $M_{\rm rec}(\pi^+\pi^-)$ distributions.
Here, $N^{\rm bg}_{\rm sb}$ and $N^{\rm bg}_{\rm bs}$ denote the numbers of peaking background events in the $E^{*}(\gamma)$ and $M_{\rm rec}(\pi^+\pi^-)$ distributions, respectively, and $N^{\rm bg}_{\rm bb}$ is the number of combinatorial backgrounds in both $A^0$ and $\Upsilon(1S)$ candidates. For $A^0\to\mu^+\mu^-$, similar 2D unbinned maximum-likelihood fits to the $M(\mu^+\mu^-)$ and $M_{\rm rec}(\pi^+\pi^-)$ distributions are performed.

In each 2D unbinned fit, the $A^0$ signal in the $E^*(\gamma)$ distribution is described by a Crystal Ball function~\cite{CB}, and that in the $M(\mu^+\mu^-)$ distribution by a double Gaussian function. The $\Upsilon(1S)$ signal in the $M_{\rm rec}(\pi^+\pi^-)$ distribution is described by a double Gaussian function. The values of the signal parameters are fixed to those obtained from the fits to the corresponding signal MC distributions. The background shapes are described by a polynomial function. All parameters are floated in the fits. We choose the order of the polynomial to minimize the Akaike information test~\cite{33201}, and find that the first-order polynomial for $M(\mu^+\mu^-)$ and second-order polynomials for $E^*(\gamma)$ and $M_{\rm rec}(\pi^+\pi^-)$ are suitable. The fitting step is approximately half of the resolution in $E^*(\gamma)$ or $M(\mu^+\mu^-)$, resulting in total of 724 and 2671 points for $A^0\to\tau^+\tau^-$ and $A^0\to\mu^+\mu^-$, respectively. From the $\tau^+\tau^-(\mu^+\mu^-)$ threshold (3.6 (0.22) GeV/$c^2$) to 9.2 GeV/$c^2$, the resolution of the $E^*(\gamma)$ distribution decreases from 5.5 MeV to 0.5 MeV, and the mass resolution of the $M(\mu^+\mu^-)$ distribution increases from 1.4 MeV/$c^2$ to 10.0 MeV/$c^2$. For each 2D unbinned fit in $A^0\to \mu^+\mu^-$ ($m_{A^0}$ $>$ 3.0 GeV/$c^2$) and $A^0\to \tau^+\tau^-$, the fitting range covers a $\pm$10$\sigma$ region.
Since the number of selected signal candidate events in the $\mu^+\mu^-$ mode with $m_{A^0}$ $<$ 3.0 GeV/$c^2$ is small, we select the following fitting intervals for different $A^0$ masses: 2$m_\mu$ $\leq$ $M(\mu^+\mu^-)$ $\leq$ 2.2 GeV/$c^2$ for 0.22 GeV/$c^2$ $\leq$ $m_{A^0}$ $\leq$ 2.0 GeV/$c^2$, and 1.8 GeV/$c^2$ $\leq$ $M(\mu^+\mu^-)$ $\leq$ 3.2 GeV/$c^2$ for 2.0 GeV/$c^2$ $<$ $m_{A^0}$ $\leq$ 3.0 GeV/$c^2$.

\begin{figure}[htbp]
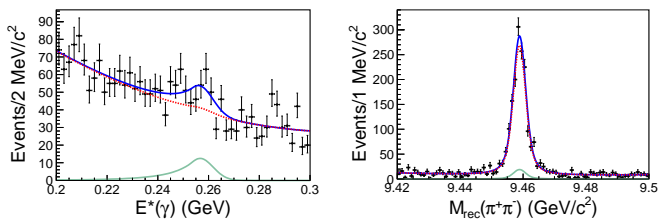

\centering
\includegraphics[width=2.78cm,angle=-90]{fig2a.epsi}
\hspace{0.2cm}
\includegraphics[width=2.78cm,angle=-90]{fig2b.epsi}
\caption{The fitted result corresponding to the maximum local significance of $3.5\sigma$ with $A^0$ mass fixed at 9.2 GeV/$c^2$ for $A^0\to\tau^+\tau^-$. The blue solid curves show the best fitted result, and the red dashed curves show the fitted total backgrounds. The green curves show the signal component.}\label{dataex1}
\end{figure}

Figures~\ref{dataex1} and \ref{dataex2} show the fitted results when the $A^0$ masses are fixed at 9.2 GeV/$c^2$ and 8.51 GeV/$c^2$ for $A^0\to\tau^+\tau^-$ and $A^0\to\mu^+\mu^-$,  respectively, where we find the maximum local signal significances for possible $A^0$ peaks. We define the local signal significance as ${\rm sign}(N_{\rm sig})\sqrt{-2\ln(\mathcal{L}_0/\mathcal{L}_{\rm max})}$~\cite{significance}, where $\mathcal{L}_0$ and $\mathcal{L}_{\rm max}$ are the maximized likelihoods without and with the $A^0$ signal, respectively. The signal yields are $116.5\pm33.4$ and $22.6\pm8.2$ with statistical significances of $3.5\sigma$ and $3.0\sigma$, respectively. The global significances are obtained to be 2.2$\sigma$ and 2.0$\sigma$ with look-elsewhere-effect included by extending the searched mass ranges to be 0.15 -- 0.4 GeV in the $E^*(\gamma)$ distribution for $A^0\to\tau^+\tau^-$ and 8.3 -- 8.7 GeV/$c^2$ in the $M(\mu^+\mu^-)$ distribution for $A^0\to\mu^+\mu^-$, respectively~\cite{C70}.
The statistical signal significances as a function of $A^0$ mass for $A^0\to\tau^+\tau^-$ and $A^0\to \mu^+\mu^-$ are shown in Figs.~\ref{ULs}(a) and~\ref{ULs}(b).
% for $A^0\to\tau^+\tau^-$ and $A^0\to\mu^+\mu^-$, respectively~\cite{C70}.

\vspace{0.2cm}

\vspace{0.2cm}
\begin{figure}[htbp]
\centering
\includegraphics[width=2.78cm,angle=-90]{fig3a.epsi}
\hspace{0.2cm}
\includegraphics[width=2.78cm,angle=-90]{fig3b.epsi}
\caption{The fitted result corresponding to the maximum local significance of $3.0\sigma$ with $A^0$ mass fixed at 8.51 GeV/$c^2$ for $A^0\to\mu^+\mu^-$. The blue solid curves show the best fitted result, and the red dashed curves show the fitted total backgrounds. The green curves show the signal component.}\label{dataex2}
\end{figure}

\vspace{0.2cm}
\begin{figure*}[htbp]
\centering
\includegraphics[width=5.5cm,angle=-90]{fig4a.epsi}
\put(-175, -10){\footnotesize $A^0\to \tau^+\tau^-$}
\put(-175, -20){\footnotesize (a)}
\put(-175, -57){\footnotesize (c)}
\put(-175, -105){\footnotesize (e)}
\hspace{0.5cm}
\includegraphics[width=5.5cm,angle=-90]{fig4b.epsi}
\put(-175, -10){\footnotesize $A^0\to \mu^+\mu^-$}
\put(-175, -40){\footnotesize (b)}
\put(-175, -57){\footnotesize (d)}
\put(-175, -105){\footnotesize (f)}
\caption{The $(a)$-$(b)$ statistical significances, $(c)$-$(d)$ upper limits at 90\% C.L. on $\BR(\Upsilon(1S)\to \gamma A^0)\BR(A^0\to \tau^+\tau^-)$ ($B_1B_2$) and $\BR(\Upsilon(1S)\to \gamma A^0)\BR(A^0\to \mu^+\mu^-)$ ($B_1B_3$), and $(e)$-$(f)$ upper limits at 90\% C.L. on $f^2_{\Upsilon(1S)}\BR(A^0\to \tau^+\tau^-)$ ($f^2_{\Upsilon(1S)}B_2$) and $f^2_{\Upsilon(1S)}\BR(A^0\to \mu^+\mu^-)$ ($f^2_{\Upsilon(1S)}B_3$) as a function of $m_{A^0}$. The blue curves show the Belle results, and the red curves show the BaBar results~\cite{071102,031102}.}\label{ULs}
\end{figure*}

The sources of systematic uncertainties in the measurements of upper limits on $\BR(\Upsilon(1S)\to \gamma A^0)\BR(A^0\to \tau^+\tau^-/\mu^+\mu^-)$ include detection efficiency, MC statistics, trigger simulation, branching fractions of intermediate states, signal parameterization, background parameterization, and total number of $\Upsilon(2S)$ events. The detection efficiency uncertainties include those for tracking efficiency (0.35\%/track), particle identification efficiency (1.1\%/pion, 1.2\%/electron, and 2.8\%/muon), and photon reconstruction efficiency (2.0\%/photon).
The above individual uncertainties from different $\tau^+\tau^-$ decay modes are added linearly, weighted by the product of the detection efficiency and all secondary branching fractions. Assuming these uncertainties are independent and adding them in quadrature, the final uncertainty related to the detection efficiency is 6.4\% for $A^0\to \tau^+\tau^-$. For $A^0\to \mu^+\mu^-$, the total uncertainty of detection efficiency is obtained by adding all sources in quadrature; it is 6.5\%.
The statistical uncertainty in the determination of efficiency from signal MC samples is 1.0\%.
We include uncertainties of 1.5\% and 1.3\% from trigger simulations for $A^0\to \tau^+\tau^-$ and $A^0\to \mu^+\mu^-$, respectively.
The uncertainty of 1.5\% from $\BR(\Upsilon(2S)\to \pi^+\pi^-\Upsilon(1S))$ is included~\cite{PDG}. The uncertainties of the branching fractions of $\tau$ decays can be neglected~\cite{PDG}.

Using the control sample of $\pi^0/\eta\to \gamma\gamma$, the maximum energy bias and fudge factor for the radiative photon are $1.004$ and $1.05$~\cite{323}, respectively. Thus, in the fitting to the $E^*(\gamma)$ spectrum for $A^0\to \tau^+\tau^-$, we change the central value by 0.4\% and energy resolution by 5\% for each $A^0$ mass point to recalculate the 90\% C.L. upper limit, and the difference compared to the previous result is taken as the uncertainty of signal parameterization.
For $A^0\to \mu^+\mu^-$, the systematic uncertainty in the mass resolution is estimated by comparing the upper limit when the mass resolution is changed by 10\% for each $A^0$ mass point. By comparing the upper limits in different fit ranges and using higher-order polynomial functions, the systematic uncertainty attributed to the background parameterization can be estimated.
The uncertainties on the total number of $\Upsilon(2S)$ events is 2.3\%.
All the uncertainties are summarized in Table~\ref{systematic} and, assuming all the sources are independent, summed in quadrature for the total systematic uncertainties.

\begin{table*}[htbp]
\centering
\caption{Relative systematic uncertainties (\%) in the measurements of upper limits for $A^0\to \tau^+\tau^-$ and $A^0\to \mu^+\mu^-$.}\label{systematic}
\vspace{0.2cm}
\begin{tabular}{c  c  c}
\hline\hline
Sources & ~~~~~$A^0\to \tau^+\tau^-$~~~~~ & $A^0\to \mu^+\mu^-$ \\\hline
Detection efficiency &6.4&6.5 \\
MC statistics &1.0&1.0 \\
Trigger & 1.5&1.3\\
Branching fractions &1.5&1.5 \\
Signal parameterization & $0.1-24.4$ & ~~$0.1-19.4$~~ \\
Background parameterization& $0.1-19.6$ & $0.1-17.2$ \\
Total number of $\Upsilon(2S)$ events &2.3&2.3 \\\hline
Sum & $7.2-32.2$ & $7.3-26.9$ \\\hline\hline
\end{tabular}
\end{table*}

We compute 90\% C.L. upper limits $x^{\rm UL}$ on the signal yields and the products of branching fractions by solving the equation $\int _0^{x^{\rm UL}}{\cal L}(x)dx/\int _0^{+\infty}{\cal L}(x)dx$ = 0.90, where $x$ is the assumed signal yield or product of branching fractions, and ${\cal L}(x)$ is the corresponding maximized likelihood of the fit to the assumption. To take into account systematic uncertainties, the above likelihood is convolved with a Gaussian function whose width equals the total systematic uncertainty.
The upper limits at 90\% C.L. on the product branching fractions of $\Upsilon(1S)\to \gamma A^0$ and $A^0\to \tau^+\tau^-/\mu^+\mu^-$ are calculated using
\begin{equation} \label{eq:add}
\begin{aligned}
\BR^{\rm UL}(\Upsilon(1S)\to \gamma A^0)\BR(A^0\to \tau^+\tau^-/\mu^+\mu^-) =
\frac{N^{\rm UL}}{N^{\rm total}_{\Upsilon(2S)}\times\varepsilon},
\end{aligned}
\end{equation}
where $N^{\rm UL}$ is the upper limit at 90\% C.L. on the signal yield, $N^{\rm total}_{\Upsilon(2S)}$ = 1.58$\times10^{8}$ is the number of $\Upsilon(2S)$ events, and $\varepsilon$ is the reconstruction efficiency with the branching fractions of $\Upsilon(2S)\to \pi^+\pi^-\Upsilon(1S)$ and $\tau$ decays included. For $A^0\to \tau^+\tau^-$, the reconstruction efficiency decreases from 2.1\% to 0.7\% with the increased $A^0$ mass, and for
$A^0\to \mu^+\mu^-$ the reconstruction efficiency decreases from 4.7\% to 0.6\% in the studied mass range from the $\mu^+\mu^-$ threshold to 9.2 GeV/$c^2$.

The upper limits at 90\% C.L. on the product branching fractions of $\Upsilon(1S)\to \gamma A^0$ and $A^0\to \tau^+\tau^-/\mu^+\mu^-$ are shown by the blue curves in Figs.~\ref{ULs}(c) and~\ref{ULs}(d), where the B$_1$, B$_2$, and B$_3$ represent $\BR(\Upsilon(1S)\to \gamma A^0)$, $\BR(A^0\to \tau^+\tau^-)$, and $\BR(A^0\to \mu^+\mu^-)$, respectively. Note that the systematic uncertainties have been taken into account. The corresponding results from BaBar~\cite{071102} are also shown by the red curves. For $A^0\to \tau^+\tau^-$, in most $A^0$ mass points, our limits are lower than those from BaBar~\cite{071102}. The most stringent upper limit can reach 4 $\times~10^{-6}$ from Belle. While from BaBar, the typical upper limit is at the level of $10^{-5}$. More stringent constraints on $A^0\to \tau^+\tau^-$ production in radiative $\Upsilon(1S)$ decays are given. For $A^0\to \mu^+\mu^-$, the upper limits at Belle are almost at the same level as those from BaBar~\cite{031102}.

The upper limit at 90\% C.L. on the product branching fractions can be converted to the Yukawa coupling $f_{\Upsilon(1S)}$ directly via~\cite{1304,1373,223}
\begin{equation}\label{fY}
\frac{\BR(\Upsilon(1S)\to \gamma A^0)}{\BR(\Upsilon(1S)\to \ell^+\ell^-)} = \frac{f^2_{\Upsilon(1S)}}{\sqrt{2}\pi\alpha}(1-\frac{m^2_{A^0}}{m^2_{\Upsilon(1S)}}),
\end{equation}
where $\ell$ = e or $\mu$ and $\alpha$ is the fine structure constant. The upper limits at 90\% C.L. on the $f^2_{\Upsilon(1S)}\BR(A^0\to\tau^+\tau^-/\mu^+\mu^-)$ as a function of $A^0$ mass are shown by blue curves in Figs.~\ref{ULs}(e) and~\ref{ULs}(f). The results from BaBar~\cite{071102} are also shown by red curves.

\begin{figure}[htbp]
\vspace{0.5cm}
\centering
\includegraphics[width=6cm,angle=-90]{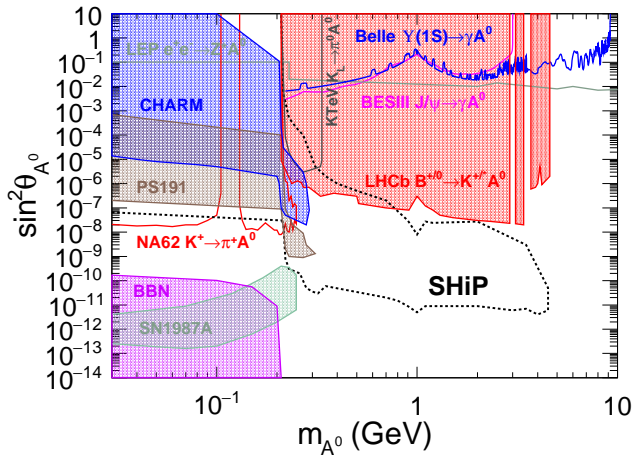}
\caption{The surviving parameter space on the plane of ${\rm sin}\theta_{A^0}$ and $m_{A^0}$. The constraints from LEP~\cite{454} (direct production of Higgs), BESIII~\cite{2109.12625} ($J/\psi$ decay), Belle ($\Upsilon(1S)$ decay), LHCb~\cite{161802,071101} ($B^{+/0}$ decay), NA62~\cite{058,093,201} ($K^+$ decay), KTeV~\cite{021805,112004,015018} ($K_L$ decay), CHARM~\cite{015018,123,010,506,124201} (beam dump), PS191~\cite{136524} (beam dump), SN1987A~\cite{003}, BBN~\cite{075033}, and the prospect of future SHiP~\cite{015018,124201} (beam dump) are shown.}\label{diff}
\end{figure}

The limit on the $A^0$ production in $\Upsilon(1S)$ radiative decays is related to the mixing angle (${\rm sin}\theta_{A^0}$), which can be compared with those from other experiments. The mixing angle is defined as~\cite{03553}
\begin{equation} \label{eq:add2}
\begin{aligned}
\frac{\BR(\Upsilon(1S)\to \gamma A^0)\BR(A^0\to {\rm hadrons})}{\BR(\Upsilon(1S)\to \ell^+\ell^-)} = \\
{\rm sin^2}\theta_{A^0}\frac{G_Fm^2_b}{\sqrt{2}\pi\alpha}\sqrt{(1-\frac{m^2_{A^0}}{m^2_{\Upsilon(1S)}})},
\end{aligned}
\end{equation}
where $G_F$ is the Fermi constant and $m_b$ is the mass of bottom quark~\cite{PDG}. When the mass of $A^0$ is smaller than $\tau^+\tau^-$ threshold, upper limits from $A^0\to \mu^+\mu^-$ are used to calculate the ${\rm sin}\theta_{A^0}$; on the contrary, upper limits from $A^0\to \tau^+\tau^-$ are used. The ratios of $\BR(A^0\to \mu^+\mu^-)/\BR(A^0\to {\rm hadrons})$ and $\BR(A^0\to \tau^+\tau^-)/\BR(A^0\to {\rm hadrons})$ are taken from Ref.~\cite{015018}; they are changed from 0.08 to 0.28 and 0.7 to 1.0 for $A^0\to \mu^+\mu^-$ and $A^0\to \tau^+\tau^-$, respectively.~The surviving parameter space on the plane of ${\rm sin}\theta_{A^0}$ and $m_{A^0}$ (the same as $m_{\phi}$ and $m_{H}$ in Refs.~\cite{015018} and \cite{03553}) from different processes are shown in Fig.~\ref{diff}.

%\vspace{0.5cm}

To conclude, we have searched for the light $CP$-odd Higgs boson in $\Upsilon(1S)\to \gamma A^0$ with $\Upsilon(2S)\to \pi^+ \pi^- \Upsilon(1S)$ tagging method using the largest data sample of $\Upsilon(2S)$ at Belle. The upper limits at 90\% C.L. on the product branching fractions for $\Upsilon(1S)\to \gamma A^0$ and $A^0\to \tau^+\tau^-/\mu^+\mu^-$ are set. In comparisons with previous studies~\cite{151802,071102,031102}, our results can further constrain the parameter space in NMSSM models~\cite{041801,051105} for $\Upsilon(1S)\to \gamma A^0(\to \tau^+\tau^-)$ and have the same restrictions for $\Upsilon(1S)\to \gamma A^0(\to \mu^+\mu^-)$.
Our limits are applicable to any light scalar or pseduo-scalar boson and dark matter, which arises in various extensions of SM.
We have used the branching fraction limits to set limits on the Yukawa coupling $f_{\Upsilon(1S)}$ and mixing angle ${\rm sin}\theta_{A^0}$. For the latter, different processes from diffferenct experiments are compared to it.

%-------- Short version, if necessary ----------------------------
%-------- * * * * * * * * * * * * * * * * * * * * *  -------------
%-------- NOTE! PRL NO LONGER REQUIRES SHORT VERSION -------------
%-------- * * * * * * * * * * * * * * * * * * * * *  -------------

We thank the KEKB group for excellent operation of the
accelerator; the KEK cryogenics group for efficient solenoid
operations; and the KEK computer group, the NII, and
PNNL/EMSL for valuable computing and SINET5 network support.
We acknowledge support from MEXT, JSPS and Nagoya's TLPRC (Japan);	
ARC (Australia); FWF (Austria); the
National Natural Science Foundation of China under
Contracts No.~12005040, No. 11575017, No. 11761141009, No. 11975076, No. 12042509, No. 12135005, No. 12161141008;
MSMT (Czechia);
ERC Advanced Grant 884719 and Starting Grant 947006 (European Union);
CZF, DFG, EXC153, and VS (Germany);
DAE (Project Identification No. RTI 4002) and DST (India); INFN (Italy);
MOE, MSIP, NRF, RSRI, FLRFAS project, GSDC of KISTI and KREONET/GLORIAD (Korea);
MNiSW and NCN (Poland); MSHE, Agreement No. 14.W03.31.0026, and HSE UBRC (Russia);
University of Tabuk (Saudi Arabia); ARRS Grants J1-9124 and P1-0135 (Slovenia);
IKERBASQUE (Spain);
SNSF (Switzerland); MOE and MOST (Taiwan); and DOE and NSF (USA).

\end{document}